\documentclass[fleqn,twoside]{article}
\usepackage{espcrc2}
\usepackage{macros,cite,amssymb}

\usepackage{graphicx}
\usepackage[figuresright]{rotating}

\newcommand{\AmS}{{\protect\the\textfont2
  A\kern-.1667em\lower.5ex\hbox{M}\kern-.125emS}}

\title{
{
\vspace{-4.0cm} \normalsize \hfill
\parbox{38mm}{DESY 03-154\\CERN-TH/2003-228\\HU-EP-03/70\\SFB/CPP-03-44\\KEK-CP-146
\\September 2003}
}\\[15mm]
\vspace{-0.0cm}
Large cutoff effects of dynamical Wilson fermions
\thanks{Talk~given~by~R.~Sommer.~Work supported in part by the 
European Community's Human Potential Programme, contract 
HPRN-CT-2000-00145, Hadrons/Lattice QCD and by the DFG in the SFB/TR 09
and the GK 271.}}
\author{R. Sommer\address[Zeu]{DESY Zeuthen, Platanenallee 6, 15738
 Zeuthen, Germany}\address[Cern]{CERN--TH, CH--1211 Geneva 23, 
Switzerland}, 
S.~Aoki\address[Utsu]{Institute of Physics, University of Tsukuba, Tsukuba, 
Ibaraki 305-8571, Japan},
M.~Della Morte\addressmark[Zeu],
R.~Hoffmann\address[HU]{Institut f\"ur Physik, Humboldt--Universit\"at zu 
Berlin, Newtonstr. 15, 12489 Berlin, Germany},
T.~Kaneko\address[KEK]{High Energy Accelerator Research Organization (KEK),
Tsukuba, Ibaraki  305-0801, Japan},
F.~Knechtli\addressmark[HU],
J.~Rolf\addressmark[HU], 
I.~Wetzorke\address[NIC]{NIC/DESY Zeuthen, Platanenallee 6, 15738
  Zeuthen, Germany} 
and U.~Wolff\addressmark[HU] 
\hspace{0.5cm}(ALPHA, CP-PACS and JLQCD Collaborations)}

\begin{document}

\begin{abstract}
We present and discuss results for cutoff effects in the PCAC masses and
the mass dependence of $r_0$ for full QCD and various fermion
actions. Our discussion of how one computes
mass dependences -- here of $r_0$ -- is also relevant for
comparisons with chiral perturbation theory. 
\vspace{1pc}
\end{abstract}

\maketitle

\section{INTRODUCTION}
\vspace{-0.1cm}
In the quenched approximation, 
large cutoff-effects were observed with the original Wilson action. 
In particular, they appeared
as a strong dependence of the PCAC-mass, 
$m=\frac12\langle \beta |\partial_\mu A_\mu |\alpha \rangle \,/\, 
\langle \beta | P |\alpha \rangle$, on the 
external states, $|\alpha\rangle,|\beta\rangle$, a dependence which 
has to be absent in the continuum limit \cite{impr:lett}.
These $a$-effects could be reduced to a tolerable level by non-perturbative 
$\Oa$-improvement \cite{impr:pap3}. The necessary improvement coefficient
$\csw$ is now known also for the Wilson gauge action 
with $\nf=2$ flavors of dynamical Wilson fermions 
(action ``W/SW'') \cite{impr:csw_nf2}
and with the Iwasaki gauge action with $\nf=3$ (I/SW)
\cite{impr:csw_iwas_nf3}.
It was pointed out already in \cite{impr:csw_nf2} that $\Oasq$ effects may 
be sizeable at the lowest value of $\beta$ considered, which roughly
corresponds to $a\approx0.1\,\fm$. Here we report on  
$\Oasq$ effects with dynamical fermions, for  
W/SW and I/SW actions and for comparison we consider also the original
Wilson (W/W) and Kogut Susskind (W/KS) actions. 
  
\section{PCAC MASSES}
\vspace{-0.1cm}
A family of external states, $|\alpha\rangle,|\beta\rangle$, 
can be realized in the Schr\"odinger Functional (SF) \cite{impr:lett,impr:pap3}.
The (bare) PCAC-mass, 
\bes \label{e_masses} \\[-5ex] \nonumber
  m = 
  \left.
  {\tilde{\partial_0} \fa(x_0) + \ca a \partial_0^* \partial_0 \fp(x_0)
   \over
   2 \fp(x_0) }\right|_{x_0=L/2} \,, 
\ees
then depends on the resolution $L/a$ of the $L^3 \times L$ space-time, on  
the spatial periodicity angle $\theta$ of the fermion fields
and the Dirichlet boundary conditions for the gauge fields.
In \eq{e_masses} the standard SF correlation functions
of the axial vector and the pseudoscalar density $\fa,\fp$ enter. 
Chiral symmetry predicts that $m$ is (apart from cutoff effects)
independent of $L/a$ as  
well as of the other parameters, as long as
the gauge coupling and the bare quark mass are kept constant.
Denoting by $m_{\infty}$ the PCAC mass computed in large volume
with $|\alpha\rangle$ the $\vecp=0$ one $\pi$ state 
and $|\beta\rangle=|0\rangle$,
one expects in particular
\bes
  \Delta m \equiv m - m_{\infty} = \Oasq
\ees
in the $\Oa$ improved cases. For W/W this difference is instead $\Oa$.
\begin{table}[t]
\centering
\begin{tabular}{cccc}
\hline \\[-2.0ex]
 action & $\beta$	& $\nf$ & 
	$\Delta m [\MeV] \, \times{a^{-1}\over 2\, \GeV}$ \\[1.0ex]
\hline \\[-2.0ex]
  W/SW	& 6.0	&	0 	&	  12(1) \\[1.0ex]
  W/SW	& 5.2	&	2	&	  28(1) \\
  W/W   & 5.5	&	2	&	  96(4) \\
  I/SW	& 2.2	&	2 	&	$-$2(4) \\[1.0ex]
\hline \\[-2.0ex]
\end{tabular}
\caption{{\small
$\Delta m$ at $a^{-1}\approx 2\GeV$, $L/a=8$.
Values for $m_{\infty}$ have been taken from 
\cite{JLQCD:nf2b52,TXL,hadr:CPPACS}. Except for $\nf=0$, the
1-loop expressions for the improvement coefficient $\ca$ are used.
For the first three lines we have $\theta=1/2$ and no background field
in the SF, while for the last line, $\theta=0$ and the background field of 
\cite{impr:pap3} was chosen.
}}\label{t_deltam} \vspace{-0.2cm}
\end{table}
Indeed, after $\Oa$ improvement
$\Delta m$ is not too large for $a\approx (2\GeV)^{-1} \approx 0.1\fm$
in the quenched approximation. However for $\nf=2$ it is more than a 
factor two larger (\tab{t_deltam}) and, although not shown here, 
this roughly 
holds also for other such mass differences. Considering several
other differences, 
we convinced ourselves that the 
only perturbatively known value for $\ca$ is not the origin of the large cutoff
effects seen for W/SW and $\nf=2$. 

We then considered the Iwasaki gauge action 
together with the values of $\csw$ used in \cite{hadr:CPPACS}. 
We found a {\em much} smaller $\Delta m$, see \tab{t_deltam}.

\section{MASS DEPENDENCE OF $r_0$}
In lattice gauge theory computations, 
the dependence of $r_0$ on the mass(es) of the dynamical quarks 
has been neglected so far. However, in contrast to the $q\bar{q}$ force at
very short distance $r$ where only an effect of relative size 
$\rmO(\alpha_{\rm s}m^2r^2)$ is present,
for the force at $r\approx 0.5\,\fm$ this dependence is not 
obviously that small.  It should be computed
by lattice gauge theory. Due to spontaneous chiral symmetry breaking,
a linear term in $m$ is expected for small quark masses.

In order to compute the mass dependence on the lattice, one must
not introduce spurious mass dependences through renormalization.
A mass independent renormalization scheme has to be chosen. In perturbation 
theory an example is provided by the lattice minimal subtraction scheme
with renormalized coupling
\bes \label{e_glat}
  \gbar^2_{\lat}(\mu) &=& Z_g(\gtilde,a\mu) \, \gtilde^2 \\
	&=& \{1 -2b_0 \ln(a\mu)  \gtilde^2 + \ldots \} \, \gtilde^2 \,.
 	\nonumber
\ees
The above equation shows  
that the lattice spacing is fixed by $\gtilde$, 
independently of the quark mass. Apart from
a small $\Oa$-correction in $\Oa$-improved QCD, which is
due to \cite{impr:pap1}
\bes
  \gtilde^2 = (1+\bg a\mq)g_0^2\,,\quad \bg=\rmO(g_0^2)\,,
\ees
the lattice spacing is determined by $g_0$ alone
in a mass-independent renormalization scheme. 
We note in passing that it is also mandatory to choose
such a scheme in a comparison of lattice QCD results with
chiral perturbation theory predictions.

To define the mass-dependence, we first introduce a
reference value $\mref$ of
the quark mass via $r_0^2\mp^2|_{m=\mref} = K$, which we later
put to $K=3$ in our numerical evaluation. Other choices are possible.
The physical mass dependence 
of $r_0$ may then be described 
by the function $\rho_{\rm cont}(x) = {r_0(m) / r_0(\mref)}$ where 
$x=m/\mref$ is finite since in the corresponding renormalized ratio
the common $Z$-factor (defined in the massless scheme) cancels.
A precise definition of $\rho_{\rm cont}$ is provided by
($\hat{r}_0=r_0/a$)
\bes
  \rho_{\rm cont}(x) = \lim_{1/\hat{r}_0(\mref) \to 0} \,\,
  \rho(x,1/\hat{r}_0(\mref))\,, \\
  \rho(x,1/\hat{r}_0(\mref)) = \left.{\hat{r}_0(m) \over \hat{r}_0(\mref)} \right|_{\gtilde \;{\rm fixed}} \,.
\ees 
As a first step we try to understand whether 
the mass dependence is a large effect and quantify
scaling violations. Since we wanted to use
results from the literature, it was easier to consider
\bes \label{e_R}
  R\left(x,\frac{1}{ \hat{r}_0(\mref)}\right) = 
  \left. {\hat{r}_0(m) \over \hat{r}_0(\mref)} \right|_{\gtilde}, \; 
  x = {r_0^2\mp^2|_m
	\over
       r_0^2\mp^2|_{\mref}	} \nonumber 
\ees 
which agrees with $\rho$ at small quark masses and lattice spacings
due to $m \sim \mp^2$.
In practice only $g_0$ could be kept constant in our evaluation of $R$,
assuming that the $\bg a\mq$ term is negligible 
($\bg \approx 0.02 + \rmO(g_0^4)$ for W/SW \cite{impr:pap1}).
It turns out that $R$ 
is well represented by the Taylor expansion
\bes
 R(x,1/\hat{r}_0(\mref)) = 1 + s(1/\hat{r}_0(\mref))\, (x-1)
\ees
for the data considered \cite{hadr:CPPACS,JLQCD:nf2b52,TXL,milc}, namely 
\bes
  \mbox{I/SW} && x\in[0.4,2.0]\,, \quad \hat{r}_0(\mref) \in [1.8,5.3] \nonumber \\
  \mbox{W/W}  && x\in[0.8,1.6]\,, \quad \hat{r}_0(\mref) \approx 4.6 \nonumber \\
  \mbox{W/SW}  && x\in[0.6,2.0]\,, \quad \hat{r}_0(\mref) \approx 4.8 \nonumber\\ 
  \mbox{W/KS}  && x\in[0.3,2.0]\,, \quad \hat{r}_0(\mref) \in [1.8,4.9] \nonumber 
\ees
Our figure shows the available results for the slope $s$. Both the 
I/SW numbers and the W/KS ones point to a very small
$|s|$ in the continuum limit with a rather strong $a$-dependence
and unclear mutual consistency.
The two other actions appear to show even larger cutoff effects
at $a\approx 0.1\,\fm$.

\begin{figure}[t]
\vspace{-0.0cm}
\centerline{\includegraphics[width=18pc]{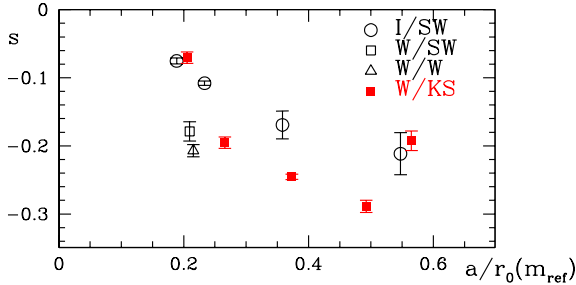}}
\vspace{-0.8cm}
\caption{Lattice
spacing dependence of slopes $s$.
}
\label{f_s}
\vspace{-0.2cm}
\end{figure}
%
\section{DISCUSSION}
\vspace{-0.1cm}
Our results are not easy to interpret, apart from the
general observation that dynamical fermions may introduce 
significant cutoff effects at $a \approx 0.1\,\fm$. 
We recall some other findings.
\bi
  \vspace{-1mm}\item	UKQCD found a very small and strongly
	mass dependent $0^{++}$ glueball mass 
	at $a\approx 0.1\,\fm$ with $\nf=2$ W/SW \cite{nf2:glueball1}. 	
  \vspace{-2mm}\item	For the same action and parameters,
	the auto-correlation times increase as the quark mass 
	is increased.
  \vspace{-2mm}\item	The pure SU(3) gauge theory has
	a phase transition in the $\beta_{\rm F},\beta_{\rm A}>0$
	half-plane not very far from $a\approx 0.1\,\fm$.
  \vspace{-2mm}\item	The JLQCD collaboration found a phase transition with
	$\nf=3$ W/SW \cite{impr:CPPACS_nf3}, preventing simulations
	unless $a$ is quite small.
\ei
These hints lead us to the conjecture that for $\nf=2$, W/SW, 
$a\approx 0.1\,\fm$, one may be close to a phase transition and
thus suffer from cutoff effects which are not necessarily described
by the Symanzik expansion. Concerning the PCAC masses (see section 2), 
I/SW is much better.
However, we are lacking an understanding why this is so and 
our figure shows significant $a$-effects in the mass dependence of $r_0$
for I/SW. 

At the present time we draw two {\bf conclusions}.\\
1) The issue of what is a good action for full lattice QCD 
deserves more
attention and one should investigate scaling properties of actions which 
are used to compute physical observables. \\
2) The mass dependence of $r_0$ and in particular the lattice artifacts 
contained in this mass dependence has to be understood before drawing  
conclusions on the chiral behavior of quantities such as $\fpi$ and $\m_\pi^2/m$.
Section 3 outlines what needs to be done but our numerical investigation
should be considered only as a first step.

{\bf Acknowledgements}. We thank B. Orth for communicating
the PCAC mass (W/W action) 
\cite{TXL}, S. Gottlieb for 
W/KS data on $r_0$  \cite{milc} and S. Hashimoto for fruitful discussions.
\vspace{-0.17cm}


\begin{thebibliography}{10}
\vspace{-0.1cm}

\bibitem{impr:lett}
K. Jansen et~al.,
\newblock Phys. Lett. B372 (1996) 275.

\bibitem{impr:pap3}
M. {L\"uscher} et~al.,
\newblock Nucl. Phys. B491 (1997) 323, hep-lat/9609035.

\bibitem{impr:csw_nf2}
ALPHA, K. Jansen and R. Sommer,
\newblock Nucl. Phys. B530 (1998) 185, hep-lat/9803017.

\bibitem{impr:csw_iwas_nf3}
CP-PACS and JLQCD collaborations, S. Aoki et~al.,
\newblock Nucl. Phys. Proc. Suppl. 119 (2003) 433, hep-lat/0211034.

\bibitem{pot:r0}
R. Sommer,
\newblock Nucl. Phys. B411 (1994) 839.

\bibitem{JLQCD:nf2b52}
JLQCD, S.~Aoki {\it et al.},
Phys.\ Rev.\ D{68} (2003) 054502,
hep-lat/0212039.

\bibitem{TXL}
T$\chi$L, N. Eicker et~al.,
\newblock Phys. Rev. D59 (1999) 014509, hep-lat/9806027.

\bibitem{hadr:CPPACS}
CP-PACS, A. Ali Khan et al.,
\newblock Phys. Rev.
      D65 (2002) 054505, hep-lat/0105015.

\bibitem{impr:pap1}
M. {L\"uscher} et~al.,
\newblock Nucl. Phys. B478 (1996) 365, hep-lat/9605038.

\bibitem{milc}
MILC, S.~Tamhankar and S.~Gottlieb, 
Nucl. Phys. Proc. Suppl., 83-84 (2000) 212 and
private communication by S.~Gottlieb.

\bibitem{nf2:glueball1}
UKQCD, A. Hart and M. Teper,
\newblock Phys. Rev. D65 (2002) 034502, hep-lat/0108022.

\bibitem{impr:CPPACS_nf3}
JLQCD, S. Aoki et~al.,
\newblock Nucl. Phys. Proc. Suppl. 106 (2002) 263, hep-lat/0110088.

\end{thebibliography}
\end{document}